# Location-Enhanced Authenticated Key Exchange

Marcos Portnoi   Chien-Chung Shen

Department of Computer and Information Sciences, University of Delaware, U.S.A.

{mportnoi,cshen}@udel.edu

*Abstract*—We introduce LOCATHE (Location-Enhanced Authenticated Key Exchange), a generic protocol that pools location, user attributes, access policy and desired services into a multi-factor authentication, allowing two peers to establish a secure, encrypted session and perform mutual authentication with pre-shared keys, passwords and other authentication factors. LOCATHE contributes to: (1) forward secrecy through ephemeral session keys; (2) security through zero-knowledge password proofs (ZKPP), such that no passwords can be learned from the exchange; (3) the ability to use not only location, but also multiple authentication factors from a user to a service; (4) providing a two-tiered privacy authentication scheme, in which a user may be authenticated either based on her attributes (hiding her unique identification), or with a full individual authentication; (5) employing the expressiveness and flexibility of Decentralized or Multi-Authority Ciphertext-Policy Attribute-Based Encryption, allowing multiple service providers to control their respective key generation and attributes.

*Index Terms*—protocol; authenticated key exchange; attribute-based encryption; security; location awareness; authentication; Bluetooth low energy

## I. INTRODUCTION

The most common form of user authentication to devices and services is the typical username/password pair. Although omnipresent, this pair might not be the most secure method of authentication, as it must strike a balance between "strong" security (i.e., having a long, random password) and not becoming excruciating to the user (i.e., allowing the user to remember the password and keeping the process of entering or typing it into the device at least moderately painless). This balance is however not trivial, in particular when a user is registered to dozens of websites/applications, and mobile devices, equipped with touch screens, do not yet offer the best experience when entering complex, random, long passwords [1].

Multi-factor authentication (MFA) is a mechanism that improves the security of passwords by combining more pieces of authentication, generally shared through diverse channels. Examples of MFA include token authenticators, SMS codes, and PINs. The combination may allow the user to manage simpler bits of authentication without loss of security. In [2], location was propositioned as a form of authentication, considering that, while utilizing a mobile device to interact with some service, the user is physically present at a "legitimate" place. If the user is indeed present at that expected location, then this information can be utilized as an authentication factor to services, i.e., in-store payment, check-ins, system access (log-on), etc.

Here, we propose LOCATHE (pronounced "locate"), a Location-Enhanced Authenticated Key Exchange, a generic protocol that combines location, user attributes, access policy and desired services as multi-factor authentication factors to allow two parties to establish an encrypted, secure session and further perform mutual authentication with pre-shared keys, passwords and other authentication factors. The combination of many factors may allow the user to manage simpler bits of authentication without loss of security. LOCATHE significantly improves on [2] and offers the following features: (1) forward secrecy through ephemeral session keys, such that past exchanges cannot be compromised if pre-shared keys or passwords are; (2) security through zero-knowledge password proofs (ZKPP), such that no passwords can be learned from the exchange; (3) the ability to use not only location, but also multiple authentication factors from a user to a service, and services are also authenticated to a user; (4) providing a two-tiered privacy authentication scheme, in which a user may be authenticated based only on her attributes (such that the user cannot be uniquely identified), or with a full individual authentication; (5) employing the expressiveness and flexibility of Multi-Authority Ciphertext-Policy Attribute-Based Encryption (CP-ABE) [3] and permitting that multiple service providers control their respective key generation and attributes. LOCATHE applies perfectly to the use case depicted in [2].

In Section II, we survey related work in authenticated key exchanges and location-enabled authentication. We present the LOCATHE protocol in Section III by utilizing, as an instance of location hardware, Bluetooth Low Energy beacons. The threat model and security of LOCATHE under several attack vectors is discussed in Section IV, and we conclude and comment on future directions in Section V.

## II. RELATED WORK

Diffie-Hellman-Merkle's seminal work [4], along with Ralph Merkle's puzzles [5], not only introduced the fundaments of public-key cryptography, but also gave one of the earliest examples of key-agreement protocols based on public key encryption, although the original idea was a non-authenticated key exchange. Subsequent work builds upon the Diffie-Hellman (D-H) exchange to offer authenticated key-exchange (AKE) protocols, or the family of Encrypted Key Exchange protocols based on the design of [6]. IKEv2 [7] is part of IPsec and provides mutual authentication between peers, forward secrecy through session keys and ways to encapsulate traffic. To address the alleged complexities of IKE, Just Fast Keying (JFK) [8] claims to achieve resistance to denial-of-service attacks, efficiency and simplicity by avoiding complex negotiations and not intending to offer perfect forward secrecy, but instead a balance between forward secrecy and efficiency through forward secrecy intervals, outside of which security associations are protected. In addition, the identities of parties are not revealed to unauthorized participants. In LOCATHE, we assume that the identity of the broadcaster is public (the first message *is* a broadcast from a known serviced location), and the identity of the user is further obscured to non-parties by session encryption, and to the service itself in one of LOCATHE's modes of privacy authentication. LOCATHE always provides forward secrecy.

Password Authenticated Connection Establishment (PACE) [9] is a ZKPP protocol intended to establish a mutually authenticated, encrypted (with strong session keys) channel between two parties utilizing short passwords for machine readable travel documents. PACE has been proposed to operate with IKEv2 [10] and accepts short keys or passwords for authentication. LOCATHE supports these short methods of authentication, augmented by the location factor. Another protocol, namely Secure Remote Password (SRP) [11], offers a verifier-based, perfect-forward-secrecy authentication in which a server stores an asymmetric form of a password (the verifier), such that leakage of the server password database provides no feasible way for an attacker to derive the actual password, or utilize the verifier to authenticate. Attribute-Based Authenticated Key Exchange (AB-AKE) [12] introduces the concept of using ABE in conjunction with AKE, in which all of the multiple parties taking part in the communication must satisfy the access policy specified by the administrator (and built within the encryption itself). The authors warn, however, that their protocol does not provide forward secrecy, although they suggest constructions that may achieve that feature. In [13], an RF transmitter mounted in a wristwatch is utilized to provide continuous authentication to nearby devices through symmetric cryptography. The system architecture provides a non-anonymous and a pseudo-anonymous (or semi-anonymous) authentication, and the authors provide a security analysis by describing the behavior of the system under a number of attack models. In particular, the authors mention *mafia fraud* and *terrorist* attacks, wherein attackers attempt to deceive the system using spoofed locations. The authors acknowledge thwarting these attacks are not trivial; in LOCATHE, we present mechanisms to mitigate the possibility of some of these location-spoofing attacks.

A location-enabled authentication system is proposed in [2], by utilizing ABE-encrypted broadcast messages and token-authenticator algorithms to provide encrypted authentication using location as an additional factor. This paper offers substantial contributions in addition to that work, in which a generic protocol is delineated to provide ZKPP, forward secrecy and protection against significant attack vectors. In [14], a detailed performance analysis on running ABE-based cryptography in smartphones is studied. The authors conclude these current devices do offer reasonable performance with ABE cryptosystems.

### III. LOCATHE, LOCATION-ENHANCED AUTHENTICATED KEY EXCHANGE PROTOCOL

In this section, we describe the construction of the LOCATHE protocol and its security components. LOCATHE is a peer-to-peer protocol, building upon robust procedures from IKEv2 and PACE [7, 10], and expands the work in [2]. The authentication protocol is divided into three stages (Broadcast, Privacy Authentication, and Exchange Authentication/Long Term Key Generation) and involves pairs of messages (a request and a response) plus one initial broadcast exchanged between two parties. The usage scenario is a user (one peer) who wants to authenticate using her present location as an authentication factor to a service (the other peer) within her proximity [1]. Location is inferred through the ability of the user in decrypting an ABE-encrypted broadcast (transmitted by Bluetooth beacons) and interactions through the protocol beginning with the broadcast. The user possesses a mobile device running an application, named the user agent. A Location-Enabled Authentication Service [2] agent (which we will subsequently name Service in this paper) within the user's location runs LOCATHE and acts on behalf of third-party service providers, named Relying-Parties (RPs)[2]. In this description, the Service Agent running LOCATHE acts as the responder, identified by the letter *r*, and the user agent (which also runs the protocol) acts as the initiator, identified by the letter *i*. We adopt this convention even though the very first message is in fact an ABE-encrypted broadcast message sent by the Service Agent, but consider the user the interested party who wants to authenticate. We utilize the Elliptic-Curve Diffie-Hellman Ephemeral (ECDHE) scheme for the key exchange operations, however the protocol may be modified to allow for the use of Ephemeral Diffie-Hellman modular (EDH) exchanges. The use of ECDHE provides robust security with comparable smaller key sizes, which is advantageous to mobile devices. Moreover, in Privacy Authentication stage (to be detailed later), LOCATHE allows the user agent to select between two privacy modes or tiers. In the Tier 1 mode, the user is only identified and authenticated to the Service based on her ability to fulfill the ABE access policy, and no individual user credentials are provided to the Service. In the Tier 2 mode, the user agent proceeds with a full authentication such that the Service individually identifies her, and may thus offer an augmented set of benefits.

*A. Registration*

Prior to authenticating, a user and the Service need to exchange data such that they may know and validate each other in subsequent interactions, and agree on security parameters. This is the registration phase, wherein a user registers with the Service running LOCATHE and exchanges security parameters such as the Service's public key, the user's ABE secret key and attributes, the seed and clock for the token authenticator algorithm, base point G for ECDHE, key-derivation function (KDF) salts, and a secret shared user key (UserKey). UserKey is a high-entropy key generated by the Service (not the user password), is typically unique per user device and can also be unique per Relying Party application. This user key expires per policy requirements, e.g., within 15 days, after which the user re-registers with the Service and acquires renewed credentials. This procedure addresses the issue of key update; a user's set of secret and shared keys are valid through a time period, after which the Service may expire them in its database, requiring the user to update them with re-registration. The ABE access policy might include special attributes that are updated periodically, such that users will need to acquire those attributes every so often in order to decrypt broadcasts within the validity period. We consider, with scalability advantages, the usage of Decentralized CP-ABE [3], wherein Services may perform their own registrations and issue their ABE secret keys to users in a manner independent of other service instances, besides common reference parameters.

---

[1] We assume "proximity" as being in the same room, or the same office floor, or the same store, a delimited space that gives a person a psychological notion of belonging to the same "area."

[2] For instance, MasterCard wishes to utilize the Location-Enabled Authentication Service running LOCATHE for in-store purchases within some store. MasterCard is thus the Relying-Party payment service provider.

*B. Broadcast/ECDHE Stage*

This stage corresponds to the broadcast of an ABE encrypted message, which furnishes both the authentication information needed for the next stage, and the exchange of session ECDHE public keys (Fig. 1 [a]). For the broadcast message, the Service picks a random nonce, Nb, which is unique per location (beacon range) and per time interval, and encrypts it using ABE with a chosen access policy, resulting in the BNONCE. (We avoid using pedix characters as possible to improve visualization within the figures.)

The Service further signs the BNONCE and broadcasts the BNONCE and the signature, and may also add its certificate to the broadcast. The random nonce Nb is regenerated periodically, triggering a new signature and broadcast. The interval between broadcasts is configurable to attain a balance between proper location accuracy, overhead and security (beacon intervals in, e.g., IEEE 801.11, are typically 100 TU, in which 1 TU = 1024 microseconds). For Bluetooth LE beacons, there is a size limit for the broadcast message. Typically, Bluetooth LE devices in advertising mode may contain a data payload of up to 31 octets [15], which is generally not sufficient for an ABE encrypted message. Therefore, the broadcast phase might require that the user agent, upon detecting the broadcast, establishes a connection with the Bluetooth LE beacon device to acquire the full ABE broadcast. As the user agent intercepts the ABE broadcast, it now becomes the initiator (i). The user agent verifies the BNONCE signature and, if the broadcaster is hence validated, attempts to decrypt BNONCE with the user's ABE secret key (ABE_sk). If the user has attributes that fulfill the access policy encapsulated within the broadcast BNONCE, then the user obtains Nb, defines an ECDHE value KEi and a random nonce Ni, and sends Ni and KEi to the Service. The Service responds with its own ECDHE value KEr and a random nonce Nr. Note that the values KEi and KEr are each calculated from the ECDHE multiplication of a secret random integer and the base point G. For instance, if the user agent picks a secret random integer *ki*, then $KEi = ki * G$.

*C. Tier 1 Privacy Authentication*

The Privacy Authentication stage comprises either a Tier 1 or a Tier 2 Privacy authentication. At Tier 1, both user and Service can compute the SharedSecret, KeySeed, and derive the session keys SK_e and SK_a as portrayed in Fig. 1 (b). SK_e is a secret key for encryption and SK_a, for authentication, and there is a different set of SK_e and SK_a keys for each direction of communications. Thus, the user agent (i) utilizes keys SK_ei and SK_ai when sending messages to the Service (r), and the Service utilizes its set of (r) keys accordingly. The pseudorandom functions *prf* and *prf+* are defined in [7], and SPIi and SPIr are unique connection identifiers resembling the SPI indexes in [16]. The SK_p keys are utilized in calculating the <SignedOctets> data that goes within the AUTH_TIER1 payloads. The <SignedOctets> data is based on the similar field in [10], however, in this tier the user agent may not provide its personal identification for privacy purposes. Additional keying material can be derived alongside the SK keys if necessary for alternative implementations of the protocol.

The user agent utilizes its session keys to encrypt and send its AUTH_TIER1_i to the Service (r). This AUTH_TIER1_i payload is produced from a combination of hashes of Nb, the nonces exchanged during Broadcast/ECDHE, <SignedOctets> (which includes SK_p and Nr) and the public ECDHE key KEr. This computation, which also obscures Nb, further mitigates replay and man-in-the-middle (MITM) attacks, as we analyze in later sections. The user agent may include optional requests such as desired services. Upon receiving the message, the Service computes the expected AUTH_TIER1_i and compares it with the user's. If matched, the user is authenticated in the Tier 1 privacy mode, i.e., the Service knows the user has the necessary attributes to fulfill the access policy, but does not know who the user is. The Service further computes its own AUTH_TIER1_r in a similar way, but also signs it with its PKI secret key and sends it to the user along the Service's identification and certificate, completing the Tier 1 Authentication.

*D. Tier 2 Privacy Authentication*

In this mode, the user utilizes her secret keys or password to authenticate herself to the Service (Fig. 1 [c]). The procedure at 1b was constructed such that it can be applied in substitution of Tier 1 sequences. However, in fact nothing forbids the parties from executing both Tier 1 and Tier 2 authentications.

As in the Tier 1 mode, both parties compute and derive their session keys and additional SK keys. To perform the individual authentication, the user agent retrieves the shared secret user key (UserKey) exchanged during registration and modifies it to a Spwd (Stored PassWorD) and then to a Kpwd (Key from PassWorD), according to Fig. 1 (c). The UserKey is never stored in the user device, but the Spwd version of it; Spwd is in fact calculated during the registration process. Alternatively, an entered user password can be utilized, such that it is properly converted into an appropriate UserKey through a KDF [17, 18] (the Service running LOCATHE again does not store, or know, the plaintext of a user password) and processed through Spwd and Kpwd. The Spwd/Kpwd modifications result in a long, high-entropy key that ultimately depends on the random, ephemeral sessions nonces exchanged in the Broadcast stage. The user agent next picks a random nonce *s* and encrypts it with the generated Kpwd, producing ENONCE. This specific encryption is done in a non-authenticated mode [19], i.e., the encryption does not generate a Message Authentication Code (MAC). This requirement intends to avoid offline password attacks, in which an attacker mounts the attacks by intercepting the broadcast messages and running them through a decryption oracle. This decryption oracle is simply an automated program that attempts decryption using several choices of passwords (and derived Spwd/Kpwd keys). If *s* is encrypted in the authenticated-encryption mode, then the attacker can easily verify whether the decryption was successful for each attempted password, regardless of the contents of the resulting plaintext. This is because the decryption oracle or program either accepts (in which case the secret key was found) or rejects decryption, since the MAC would be invalid for the attempted key. The user further computes GE based on *s* and the previously calculated

SharedSecret and creates a pair LSK_i and LPK_i, which are, respectively, the secret and public user keys for Tier 2 authentication. The user then encrypts (with the session keys) the ENONCE, her user ID (IDi), AUTH_TIER2_i (which is a proof of knowledge of the broadcast Nb and exchanged data so far) and LPK_i, and sends all of them to the Service. The UserKey or user password are never sent over the medium, even in encrypted form; the protocol provides a zero-knowledge password proof of it through ENONCE. The Service verifies the user ID and similarly retrieves Spwd and generates Kpwd from the associated UserKey, thus being able to decrypt ENONCE. It utilizes the obtained random nonce *s* to compute GE and its pair of secret and public keys LSK_r and PSK_r. It then encrypts and sends its ID parameters/certificate (IDr) and its public key LPK_r to the user agent, along with possible requests for additional authentication factors. When the user agent provides additional authentication factors, each of these factors should be converted into strong keys as demonstrated by the Spwd/Kpwd construction. If the retrieved *s* is different from that used by the user agent (due to a potential successful malicious attack), the resulting GE and key pairs will differ from the expected values and cause a failure in the authentication in the next stage.

*E. Exchange Authentication/Long Term Key Generation*

In this final stage (Fig. 1 [d]), the parties further authenticate themselves and the exchanges so far, and finally compute a long term shared key that can be utilized in new location authentications and potential session hand-offs, such as transferring the session from Bluetooth LE to Wi-Fi. If the Service has requested additional authentication factors from the user agent, such as extra passwords, tokens, PINs or biometric data, they are sent in this step. We utilize, as an example, a token authenticator (*tk*)-generated number. The GTK number is computed from the *tk* number and the ECDHE point GE, such that the actual *tk* is never sent from the user agent to the Service. The user agent computes AuthSharedSecret from its secret LSK_i and the received Service's public LPK_r, and then generates the authentication data AUTHi. The <SignedOctets_i> data is based on the similar field in [10]. The user agent then sends AUTHi to the Service. The Service computes GTK (since the Service shares the same seed, clock and token authenticator algorithm with the user, it can generate the same numbers at the same time), AuthSharedSecret and its AUTHr data in a similar fashion, and then verifies the received AUTHi. If this final authentication is valid, the Service proceeds to send its AUTHr to the user agent and compute the long-term secret key LongTermSecret, which may be stored and substitute UserKey in further Tier 2 authentications. LongTermSecret shall expire within 1 hour of creation, following the guidelines suggested in [10]. The user agent, upon receiving and verifying AUTHr, also computes LongTermSecret. Both the AUTH values and the LongTermSecret are tied to the ephemeral, random values created during the location-enabled authentication session.

## IV. SECURITY ANALYSIS AND THREAT MODEL

We provide in this section a descriptive analysis of the behavior of LOCATHE under different attack vectors. The threat model is constructed as a game with three players, each named (as tradition) as Alice, Bob, Mallory: the Service Agent running LOCATHE protocol (Alice) and a user (Bob) are legitimate peers who wish to establish a location-enabled authenticated communication. Mallory is a malicious attacker; she is modeled as being able to listen to, record and replay the exchanges between Alice and Bob, or communicate directly with Alice and/or Bob (thus, being passive or active) with messages of her choosing. Mallory knows no pre-shared secret keys nor passwords, and wins the game if she is able to determine confidential information exchanged between Alice and Bob with non-negligible probability, or successfully authenticate herself as either Alice or Bob.

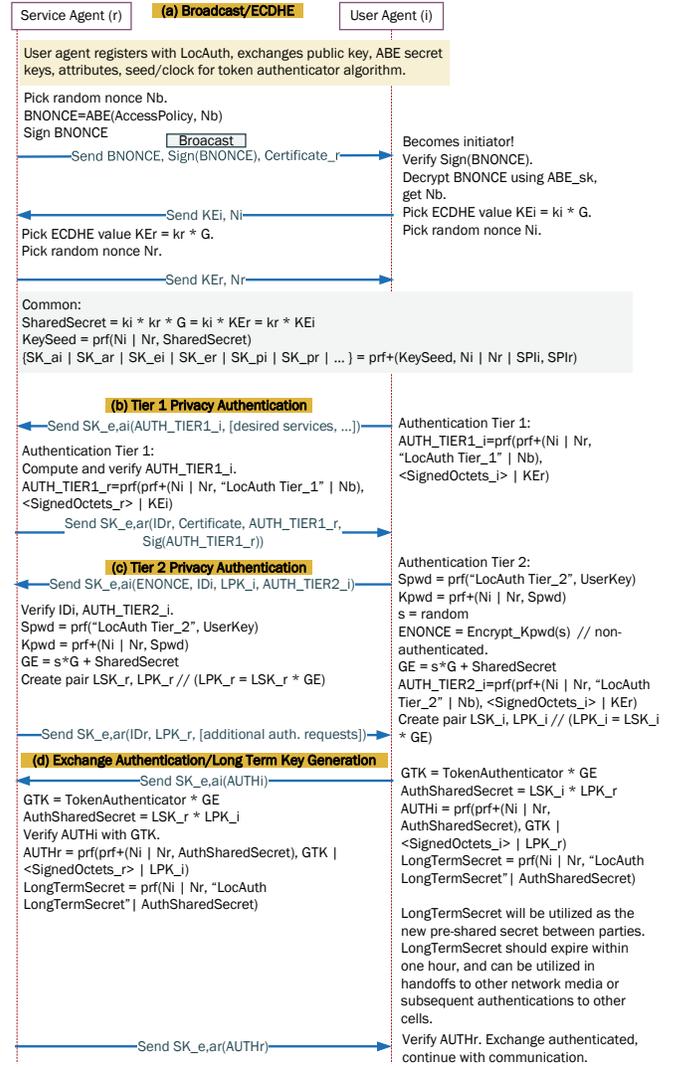

Fig. 1: The stages and request/response messages of the LOCATHE protocol.

*A. Eavesdropping*

In this attack, Mallory passively listens to the exchanges between Alice and Bob and attempts to learn confidential data encrypted by the session keys, ABE encryption, or pre-shared keys. The only information Mallory is able to read is the public ECDHE keys and public nonces, which are not encrypted. After the Broadcast stage, all exchanges are encrypted with ECDHE ephemeral keys that Mallory cannot derive as long as the assumption holds that an attacker cannot solve the Elliptic Curve Discrete Logarithm Problem. Moreover, even if Mallory successfully learns some pre-shared secret, she cannot decrypt past sessions, and even learning one session key does not provide the ability to decrypt the other direction of exchanges, giving the protocol forward secrecy.

*B. Man in the Middle Attack (MITM)*

A non-authenticated Diffie-Hellman exchange is susceptible to active attackers, i.e., attackers who are capable of intercepting the ECDHE values and substitute them with their own values. Effectively, in MITM, Mallory may intercept the initial ECDHE values send by Alice and Bob, sending her own computed ECDHE values and establishing two encrypted sessions (with dissimilar keys), one between Alice and Mallory, and one between Mallory and Bob. LOCATHE, however, utilizes authentication in different stages to repel the MITM attack, and the parties never send keys or passwords. The initial broadcast consists of the encrypted, temporary random nonce Nb with ABE, which Mallory cannot obtain without breaking the ABE encryption; this Nb is further utilized to produce authentication data in the Privacy Authentication steps that ties crucial ephemeral values exchanged between the parties. Without Nb, the resulting invalid data prevents mutual authentication. Since the actual value Nb is never sent back to the originator, but a hash-function transformation of it, Mallory cannot learn the original Nb. Finally, the encrypted Nb is signed by the Service (Alice), allowing the user agent (Bob) to reject tampering.

Moreover, in Tier 2 Privacy authentication, a shared secret key and a second-factor token authenticator validate the user agent in a ZKPP exchange with the Service. In the last stage, additional authentication fields are constructed with exchanged and derived data that cannot (but with negligible probability) be obtained, or provided, by the attacker without resulting in invalid authentication values that both Alice and Bob reject. Unless Mallory is able to compromise pre-shared and the ABE secret keys (in which case all security is lost), Mallory cannot impersonate Alice nor Bob, and LOCATHE rejects this attack.

*C. Replay Attack*

Here, Mallory listens to and record transmissions between Alice (Service) and Bob (legitimate user), and later replays her recordings in the same location. (Replays in a different location are addressed next as *wormhole attacks.*) Mallory can choose which recording and which time to replay, and she wins the game if she successfully impersonates Alice or Bob, or learns confidential information. Aforementioned mechanisms also thwart replay attacks, in particular the ephemeral, random values that are utilized to generate shared secrets, challenges and authentication payloads. Simply replaying the other party of an exchange with an active Alice or Bob has negligible probability of success, since the legitimate party must choose the exact same random values occurred in the replayed session, such that an exact reproduction of a past exchange ensues. In particular, we emphasize the initial Nb broadcast and the token authenticator in Tier 2 authentication. The Nb broadcast is unique per location and continually regenerated by Alice. Let $t$ be the time in which Alice broadcasts the ABE-encrypted message $m$; let $d$ be the time interval within which Alice considers $m$ valid; and let $\delta$ be a real number such that $\delta > 0$. We identify two sub-cases: (a) Mallory replays the Nb broadcast in time $t + d + \delta$ (outside the valid time interval); and (b) Mallory replays the broadcast in time $t' \leq t + d$ (within the valid period). In case (a), Bob will still respond to the replayed message (since it appears valid to Bob) and may complete the Broadcast/ECDHE stage with either Alice or Mallory. In the Privacy Authentication steps, Bob sends his authentication payload constructed with Nb to Alice or Mallory. The current broadcast message by Alice will contain a different random value, say Nb'. The difference will result in invalid authentication payloads from Bob, which Alice rejects. Mallory cannot successfully establish an ECDHE encrypted session with Bob, past the Broadcast step, or decrypt the session with non-negligible probability just by replaying past responses. In case (b), Bob will either (i) reply to it for the first time if he has not yet heard the original broadcast, in which case the reply will be captured by Alice who will then continue with a valid exchange with Bob, or (ii) if Bob has already replied to this same broadcast, he will simply disregard it as a duplicate. Not shown in the figures, a message counter mechanism allows the parties to identify and reject duplicates. Finally, Tier 2 authentication includes a token authenticator, which generates diverse, temporary values only known by Alice and Bob, mitigating replay attacks. If Mallory instead replays messages from Bob to Alice, it is easy to verify that the same protocol behavior occurs.

*D. Wormhole Attack*

The setup for the wormhole attack comprises two locations, *l* and *p*, wherein messages transmitted in one location cannot normally be heard at the other. Both locations have a "local" Alice, i.e., LOCATHE service beacons. Mallory listens to exchange messages between Alice and Bob at location *l*, and she has a direct network connection through a secondary channel to location *p*. Mallory wins the game if she successfully gains authentication with the Service at location *p*, by replaying the interactions or responses from Bob to the local Alice at *l*. That is, Mallory attempts to replay legitimate interactions from location *l* to gain authentication with the Service at another location *p*.

If the Service at location *p* receives messages from Bob who is interacting with the Alice beacon at location *l,* there is negligible probability that the SPI indexes and the encrypted broadcasts at both locations are exact matches. If they are not, authentication payloads will be invalid, and thus the Service at *p* rejects authentication. Alternatively, Nb may contain location or beacon ID data, facilitating matching responses to broadcast originators.

A different wormhole attack may be performed if Mallory can listen and replay messages between locations *l* and *p*. In this case, Alice is at *l*, and Bob is at *p*, and there is no "local" Alice Service at *p*. Mallory is, thus, effectively extending the range of location *l* to location *p* through her network connection. LOCATHE does not protect against this form of attack, assuming Bob has no other location-identifier hardware.

*E. Denial of Service (DoS) Attack*

Mallory performs a DoS attack by either (a) replaying previously recorded messages, (b) transmitting bogus messages, or (c) transmitting radio noise (jamming) with such transmission power and constancy to disrupt legitimate communications between Alice and Bob. As described in [2, 20], Bluetooth frequency hopping mitigates radio jamming. The use of unique SPI connection indexes and message counters assist the parties in ignoring bogus or repeated messages, although there will still be overhead in detection. Not described in LOCATHE, [7, 21] suggests the use of cookies that may assist the Service in detecting and further rejecting repeated attempts of session initiation.

V. CONCLUSION AND FUTURE WORK

We introduced LOCATHE, a Location-Enhanced Authenticated Key Exchange protocol. LOCATHE permits capturing, through Attribute-Based Encryption-encrypted broadcasts, the dynamic association of location, user's attributes, access policy and local service as multi-factor authentication such that two parties, a user and a service, can establish an encrypted, secure session and mutually authenticate with additional factors. Security features of LOCATHE include forward secrecy; zero-knowledge password proofs to avoid password or key leakages; use of strong keys; capability of accommodating multi-factor authentication; a two-tiered privacy authentication scheme, wherein a user can choose to be fully, individually authenticated, or authenticated without personal identification. The underlying Location-Enabled Authentication Service employs Multi-Authority CP-ABE, so that participating services and Relying Parties can control their own ABE secret key generation and update, and access policies. By generating a strong, long-term shared key, LOCATHE contributes a secure way for users to authenticate to different participating services with less user intervention, using the location/mobility information as a security factor. Our future work includes evaluating the performance of LOCATHE and its overhead, and investigating its security under extended attack vectors. Additional or alternative password-based authentication schemes will be analyzed, in particular verifier-based ones such as SRP. LOCATHE provides resources for key update and (implicit) revocation through periodic updates, however we plan to survey possible more efficient constructions.